\documentclass{emulateapj}

\usepackage{apjfonts}
\usepackage{natbib}
\usepackage{amssymb,latexsym}
\usepackage{graphicx}
\usepackage{appendix}
\usepackage{amsmath}
\usepackage{float}

\newlength{\colwidth}

\usepackage[usenames,dvipsnames]{xcolor}

\newcommand{\Msun}{\rm M_{\odot}}
\newcommand{\Mpcph}{\mathrm{Mpc}~h^{-1}}

\newcommand{\zreion}{z_{\mathrm{reion}}}
\newcommand{\treion}{t_{\mathrm{reion}}}
\newcommand{\taues}{\tau_{\rm es}}
\newcommand{\labs}{\lambda_{\rm abs}}
\newcommand{\Mmin}{M_{\rm min}}
\newcommand{\hii}{\textsc{Hii}\ }

\newcommand{\beq}{\begin{equation}}
\newcommand{\eeq}{\end{equation}}

\shortauthors{Li et al}
\shorttitle{Reionization Histories of Milky Way Mass Halos}

\begin{document}
\title{ Reionization Histories of Milky Way Mass Halos }
\author{Tony Y. Li$^1$, Marcelo A. Alvarez$^2$, Risa H. Wechsler$^1$, Tom Abel$^1$}
\affil{$^1$Kavli Institute for Particle Astrophysics and Cosmology;
 Physics Department, Stanford University, Stanford, CA 94305, USA\\
 SLAC National Accelerator Laboratory, Menlo Park, CA 94025, USA;\\
 tonyyli@stanford.edu, rwechsler@stanford.edu, tabel@stanford.edu\\
$^2$CITA, University of Toronto, Toronto, Ontario, Canada; malvarez@cita.utoronto.ca}
\date{\today}

\begin{abstract}
  We investigate the connection between the epoch of reionization and
  the present day universe, by examining the extended mass
  reionization histories of dark matter halos identified at $z=0$.  We
  combine an N-body dark matter simulation of a 600 Mpc volume with a
  three-dimensional, seminumerical reionization model.  This provides
  reionization redshifts for each particle, which can then be
  connected with the properties of their halos at the present time.
  We find that the vast majority of present-day halos with masses
  larger than $\sim$ few $\times 10^{11} \Msun$ reionize earlier than
  the rest of the universe.  We also find significant halo-to-halo
  diversity in mass reionization histories, and find that in realistic
  inhomogenous models, the material within a given halo is not
  expected to reionize at the same time.  In particular, the scatter
  in reionization times \emph{within} individual halos is typically
  larger than the scatter \emph{among} halos.  From our fiducial
  reionization model, we find that the typical 68\% scatter in
  reionization times within halos is $\sim 115$ Myr for $10^{12\pm
  0.25} \; \Msun$ halos, decreasing slightly to $\sim 95$ Myr for
  $10^{15\pm 0.25}\;\Msun$ halos.  We find a mild correlation between
  reionization history and environment: halos with shorter
  reionization histories are typically in more clustered environments,
  with the strongest trend on a scale of $\sim$ 20 Mpc.  Material in
  Milky Way mass halos with short reionization histories is
  preferentially reionized in relatively large \hii regions, implying
  reionization mostly by sources external to the progenitors of the
  present-day halo.  We investigate the impact on our results of
  varying the reionization model parameters, which span a range of
  reionization scenarios with varying timing and morphology.
  \end{abstract}

\maketitle

\section{Introduction}

The signs of reionization should survive to the present day universe.
Although the epoch of reionization is constrained to redshifts $z
\gtrsim 6$ by quasar absorption spectra
\citep[e.g.,][]{becker/etal:2001, fan/etal:2006} and the cosmic
microwave background \citep{komatsu/etal:2011, hinshaw/etal:2013}, the
process is likely to be temporally extended and spatially
inhomogeneous \citep{gnedin:2000, miralda-escude/etal:2000,
sokasian/etal:2001, barkana/loeb:2004, iliev/etal:2006,
trac/cen:2007}.  One also expects the process to be closely coupled to
the formation of early galaxies: on one hand, high-redshift, low-mass
galaxies may be essential sources of ionizing radiation
\citep[e.g.,][]{wise/cen:2009,bouwens/etal:2012,
kuhlen/faucher-giguere:2012, alvarez/etal:2012}. On the other hand,
reionization may photoheat gas in halos with virial temperatures
$T_{vir} \lesssim 10^4$ K, suppressing or truncating galaxy formation
in those low-mass halos
\citep{efstathiou:1992,shapiro/etal:1994,thoul/weinberg:1996,
barkana/loeb:1999,gnedin:2000b,dijkstra/etal:2004,okamoto/etal:2008}.

Indeed, this proposed suppression of galaxy formation may partially
alleviate the so-called ``missing satellites'' problem
\citep{moore/etal:1999, klypin/etal:1999, bullock/etal:2000,
kravtsov/etal:2004}, in which cold dark matter simulations predict far
more dark matter subhalos in the Local Group than observed satellite
galaxies.  Reionization provides a plausible mechanism for suppressing
the present-day faint satellite population, reducing the missing
satellite discrepancy.  Recently, the discovery of so-called
``ultra-faint dwarf'' galaxies around the Milky Way
\citep[e.g.,][]{willman/etal:2005, zucker/etal:2006,
belokurov/etal:2007} has partially closed the gap.  However, there is
also evidence that their star formation histories may have been
truncated during reionization \citep{brown/etal:2012}.  If, indeed,
reionization directly affects the present-day population of faint
satellites \citep{koposov/etal:2009, busha/etal:2010,
ocvirk/aubert:2011}, or leaves its mark in other ways, then it is
important to quantify the reionization epoch and history of the Local
Group environment \citep{weinmann/etal:2007, iliev/etal:2011}.

Self-consistently modeling the competing, simultaneous processes
involved in reionization will be a key component of a comprehensive
theory of galaxy formation.  In particular, three-dimensional
simulations are necessary to capture the patchy morphology and
evolution of the ionized \hii regions.  Unfortunately, the ideal of
direct, radiative transfer simulations of reionization, over all
scales involved, is prohibitively expensive on present-day computers.
However, insofar as we would like to model the large scale morphology
of reionized regions, approximate seminumerical schemes have shown
good agreement with radiative transfer simulations, on large scales,
at a fraction of the computational cost \citep{zahn/etal:2007,
zahn/etal:2011, santos/etal:2010, mesinger/etal:2011}. 

A number of previous studies have directly informed our focus here.
\citet{alvarez/etal:2009b} combined three-dimensional reionization
calculations with an N-body simulation to assign reionization epochs
to present-day dark matter halos.  \citet{busha/etal:2010} applied the
resulting distribution of reionization redshifts to Via Lactea II
subhalo catalogs, finding that varying the Milky Way reionization
epoch could alter the satellite population of Milky Way halos by up to
2 orders of magnitude, while assuming for simplicity that the halo
reionization epoch was instantaneous.  \cite{lunnan/etal:2012}, using
a similar method for calculating satellite luminosities, explicitly
modeled the effect of patchy reionization on the satellite population
of six Milky Way-like halos. Comparing instantaneous and inhomogeneous
reionization scenarios, they found differences of 10-20\%---both
increase and decrease---in the number of faint ($M_V \gtrsim -10$)
satellites, suggesting against a systematic correction between the two
scenarios. Additionally, they found a large halo-to-halo scatter in
satellite populations across a small sample of 6 halos, motivating the
study of a larger, statistical sample of halos.

In order to further investigate and quantify the connection between
patchy reionization and present-day halos, we have therefore chosen an
N-body simulation which includes both (1) well-resolved $z=0$ halos
down to Milky Way masses, and (2) a large, statistical sample of such
halos. In this study, we focus on calculating the \emph{extended}
reionization histories of present-day halos in order to quantify the
inhomgeneity of their reionization epochs.

\section{Modeling}

\subsection{Reionization Model} \label{sec:reionmodel}

We employ a seminumerical reionization model, which we summarize here,
based on the analytic prescription developed by
\cite{furlanetto/etal:2004} and extended to three dimensions by
\cite{zahn/etal:2007}.  We include an additional treatment of
photon absorption by Lyman-limit systems, following the implementation
of \cite{alvarez/abel:2012}; further details may be found in that 
paper.

The fundamental assumption of the model is that a given region is
completely ionized if enough photons have been emitted by enclosed
sources to ionize it.  Formally, in order to ionize the region we
require that $f_{\rm coll}$, the fraction of matter collapsed in halos
above some mass threshold $\Mmin$, satisfies the inequality 
\begin{align}
	\zeta f_{\rm coll} & \geq 1 \label{eq:zetafcoll}
\end{align}
where $f_{\rm coll}$ is the fraction of mass in collapsed objects within
the region.  The efficiency factor $\zeta$, which may be interpreted
as the effective number of ionizing photons released per collapsed atom,
encapsulates all galaxy formation and radiative transfer physics.  Via
the excursion-set formalism \citep[e.g.,][]{1991ApJ...379..440B,lacey/cole:1993}, $f_{\rm
  coll}$ in a spherical region of mass $m$ and overdensity $\delta$
may be written 
\begin{align}
	f_{\rm coll} &= \mathrm{erfc} \left\{ \frac{\delta_c(z) - \delta_m}{\sqrt{2 [\sigma^2(M_{\rm min}) - \sigma^2(m)]}} \right\}.
\end{align}

At any given position, the density field is smoothed with a spherical
real-space top hat filter, over increasing radii $R$ (or alternatively
mass scales $m$), and the earliest redshift, at any smoothing scale,
for which Equation \ref{eq:zetafcoll} is satisfied is recorded as the
reionization redshift $\zreion$ of that point. To ensure global photon
conservation, we follow the procedure also outlined in
\citet{alvarez/abel:2012}, in which the reionization redshift of each
cell is modified so that the order in which cells are ionized is
preserved, but the overall reionization history is given by
\begin{align}
	\frac{dx}{dt} &= \zeta\frac{{d\bar{f}}_{\rm coll}}{dt} \frac{\labs}{\labs + \lambda_{\rm b}(x)} 
\end{align}
where $\overline{f}_{\rm coll}$ is the mean collapsed fraction and
$\lambda_{\rm b}(x)$ is the evolving mean distance between neutral
patches -- the "bubble mean free path" \citep{alvarez/abel:2012}. 

To perform the calculation, the simulation box is uniformly subdivided
into $1400^3$ cells, and the smoothing procedure is performed around
the center of each cell, assigning it a reionization redshift. There
are exactly as many cells as simulation particles (see \S
\ref{sec:nbodysim}), linking every cell to a unique initial,
unperturbed simulation particle. Because the excursion set formalism
is fundamentally Lagrangian---i.e., it follows mass rather than volume
elements---this directly tags each particle with its own reionization
redshift $\zreion$, or alternately a lookback reionization time,
$\treion$.  Assuming that baryons trace the dark matter in our
simulation, this allows us to characterize the reionization times of
baryonic matter in present-day halos.

\begin{table}[!t]
	\centering
	\caption{Reionization Model Parameters}
	\begin{tabular}{r|c|c||c|c|c}
		$\taues$			\footnote{Electron scattering optical depth to reionization.}		& 
		$\Mmin$ [$M_\odot$]	\footnote{Minimum halo mass of ionizing sources.}					&
		$\labs$ [$\Mpcph$]	\footnote{Mean free path of residual Lyman absorption systems.}		&
		$\zeta$				\footnote{Ionizing efficiency, implicitly set by other parameters.}	&
		$z_{\rm 0.5}$		\footnote{Redshift when global reionized mass fraction is 0.5.}		&
		$z_{\rm end}$		\footnote{Redshift when global reionized mass fraction is 1.0.}		\\
		\hline
		0.06		& $10^8$		& 8			& 18.1		& 8.4		& 3.6		\\
		0.06		& $10^8$		& 32		& 14.6		& 8.3		& 4.9		\\
		0.06		& $10^8$		& 256		& 13.1		& 8.2		& 6.4		\\
		0.06		& $10^9$		& 8			& 72.9		& 8.5		& 5.3		\\
		0.06		& $10^9$		& 32		& 56.0		& 8.4		& 6.0		\\
		0.06		& $10^9$		& 256		& 50.0		& 8.3		& 7.1		\\
		\hline
		0.09		& $10^8$		& 8			& 113.0		& 11.3		& 7.7		\\
		\footnote{Bold: fiducial model.}
		\bf 0.09	& $\bf 10^8$	& \bf 32 	& \bf 90.9	& \bf 11.2	& \bf 8.5	\\
		0.09		& $10^8$		& 256		& 81.0		& 11.1		& 9.7		\\
		0.09		& $10^9$		& 8			& 1020.0	& 11.4		& 9.0		\\
		0.09		& $10^9$		& 32		& 785.0		& 11.3		& 9.5		\\
		0.09		& $10^9$		& 256		& 700.0		& 11.3		& 10.3		\\
		\hline                                                                
		0.12		& $10^8$		& 8			& 812.0		& 13.9		& 10.9		\\
		0.12		& $10^8$		& 32		& 657.0		& 13.8		& 11.6		\\
		0.12		& $10^8$		& 256		& 580.0		& 13.8		& 12.5		\\
		0.12		& $10^9$		& 8			& 1.82e4	& 14.0		& 12.0		\\
		0.12		& $10^9$		& 32		& 1.41e4	& 13.9		& 12.4		\\
		0.12		& $10^9$		& 256	 	& 1.24e4	& 13.9		& 13.0		
	\end{tabular}
	\label{tab:reionmodels}
\end{table}

Our model is parameterized by three global values:
\begin{enumerate}
	\item $\taues$, the optical depth of CMB photons to scattering from free electrons
	\item $\labs$, the comoving, spatially uniform mean free path of absorption systems
	\item $\Mmin$, the minimum halo mass of ionizing sources
\end{enumerate}
The ionizing efficiency $\zeta$ is then implicitly determined from
these parameters.  Table \ref{tab:reionmodels} lists all parameter
values explored, and we briefly discuss the effect of varying these
parameters in \S \ref{sec:varyingparams}.  For our fiducial model, we
use $\taues = 0.09$, $\labs = 32 \;\Mpcph$, $\Mmin=10^8 \;\Msun$, in
which reionization is completed by $z_{\rm end} \approx 8.5$.  This is
somewhat early relative to the $z_{\rm end} \gtrsim 6$ constraint
inferred from quasar spectra, but ensures a global reionization
history consistent with the inferred value of $\taues$ from the WMAP
7-year result, $\taues=0.088\pm 0.014$ \citep{komatsu/etal:2011} [note
that it remains consistent with subsequent WMAP 9-year ($0.081 \pm
0.012$) and Planck ($0.092 \pm 0.013$) results
\citep{bennett/etal:2012, planck1:2013}]. We have evaluated
$\taues$ here as: 
\beq
	\taues 
	= \frac{3 H_0 \Omega_b c \sigma_T}{8\pi G m_p} \int_0^\infty
	\frac{x(z)(1+z)^2(1-Y+N_{He}(z)Y/4)}{\sqrt{\Omega_m(1+z)^3+1-\Omega_m}} dz, 
\eeq
where $Y\simeq 0.24$ is the helium abundance and $N_{\rm He}(z)$ is
the number of times helium is ionized---we use $N_{\rm He}=2$ for
$z<3$ and $N_{\rm He}=1$ for $z>3$. 
 
We acknowledge the simplicity inherent in a model, such as this one,
which does not explicitly include radiative transfer.  However,
comparisons of radiative transfer and seminumerical codes have been
performed \citep[e.g.,][]{zahn/etal:2011}, in which both schemes
predict similar \hii morphologies and are in excellent agreement on
scales of $\gtrsim {\rm Mpc}$.  The seminumerical scheme does so at a
fraction of the computational cost, justifying its use when such
large-scale reionization morphology is the main concern, as it is
here.

\begin{figure}[!htbp]
		\centering
		\includegraphics[width=\colwidth]{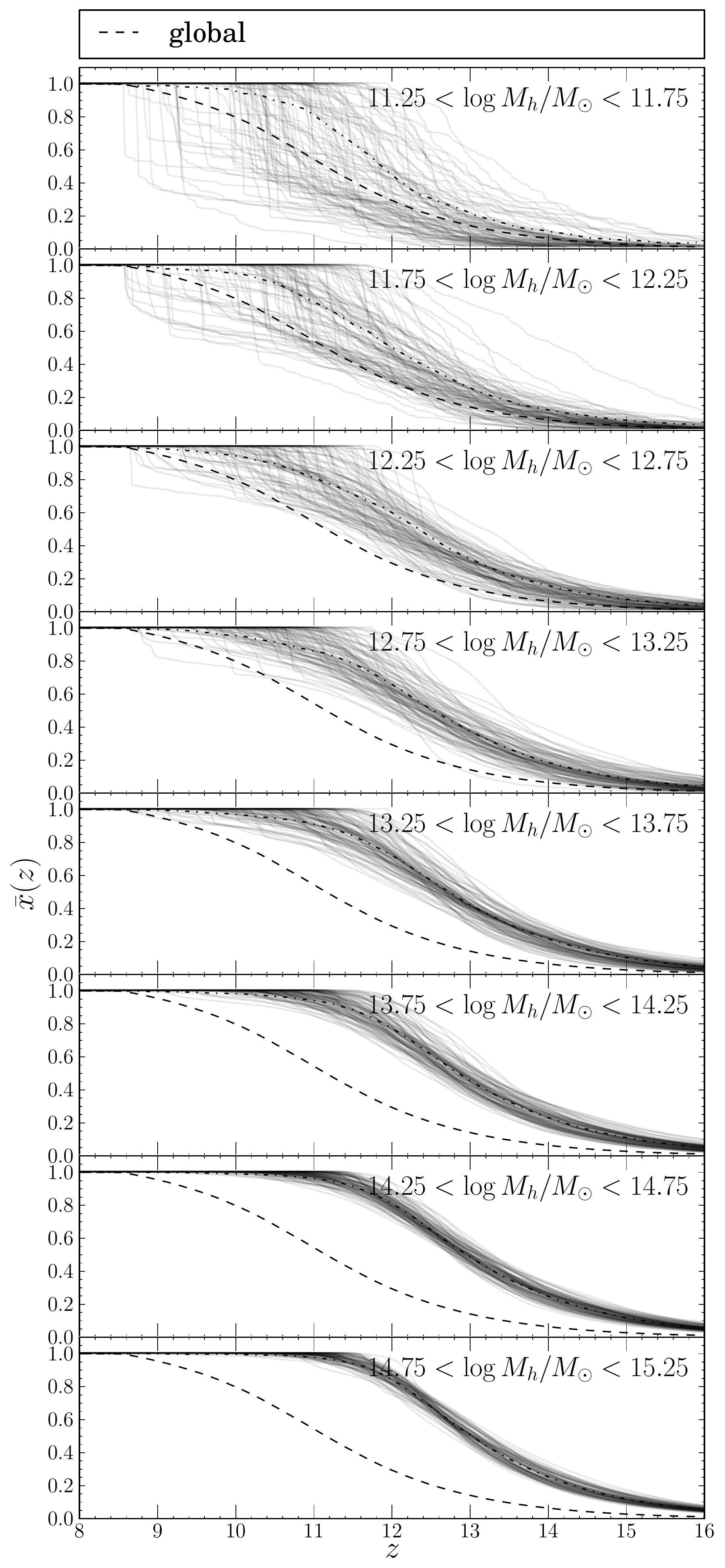} 
		\caption{Randomly-selected halo reionization histories, in
			increasing mass bins of width 0.5 dex.  The reionized
			fraction $\bar{x}(z)$ of each halo's $z=0$ mass is plotted
			as a function of redshift.  100 halos were selected in
			each mass bin.  The global reionized mass fraction ({\em
			dotted}) and the mean halo reionized fraction ({\em
			dash-dotted}) are also shown in each bin.}
		\label{fig:haloreionhistories}
\end{figure}

\begin{figure}[!t]
		\centering
		\includegraphics[width=\colwidth]{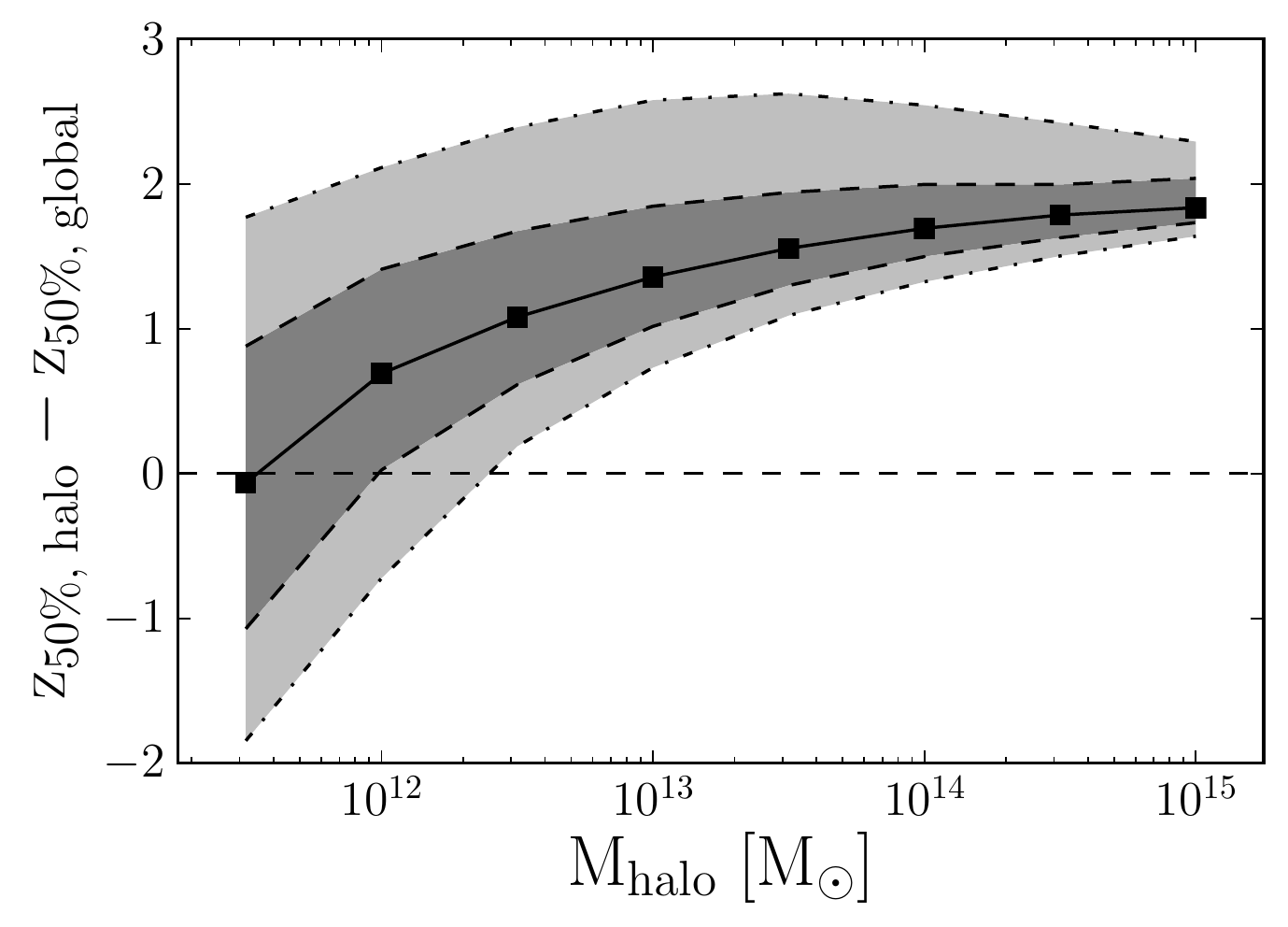} 
		\caption{Offset between halo and global half-reionized
			redshifts, as a function of $z=0$ halo mass. In each mass
			bin, the median ({\em solid black squares}) is indicated,
			along with lines tracing the 68\% ({\em dashed}) and 95\%
			({\em dash-dotted}) distributions.  A halo which lies
			above (below) the dotted line essentially reionizes before
			(after) the rest of the universe.} 
		\label{fig:halosreionizefirst}
\end{figure}

\subsection{N-Body Simulation} \label{sec:nbodysim}

Although our reionization model is calculated solely from the initial
overdensity field, we wish to connect those calculations to
present-day halos.  To that end, we use one realization of the
\emph{Consuelo} $N$-body simulations from the Large Suite of Dark
Matter Simulations\footnote{http://lss.phy.vanderbilt.edu/lasdamas}.
The simulation contains $1400^3$ dark matter particles, each with mass
$m_p = 2.67 \times 10^{9} \;M_\odot$, within a comoving 600
Mpc box with periodic boundary conditions.  Particles were
evolved forward in time with the \textsc{Gadget}-2 code
\citep{springel:2005}, and in this study we assume cosmological
parameters consistent with those of the simulation, namely $\Omega_M =
0.25$, $\Omega_\Lambda = 0.75$, $h=0.7$, $\sigma_8 = 0.8$, $n_s =
1.0$.

In the simulation, dark matter halos were identified using the
\textsc{Rockstar} phase-space halo finder \citep{behroozi/etal:2011},
which resolved $\sim 1.6 \times 10^6$ halos in the mass range $10^{11.25}
M_{\odot} \lesssim M_{\rm halo} \lesssim 10^{15.25} M_{\odot}$.  In this
study, we have not counted another $\sim 4.5 \times 10^5$ subhalos,
i.e. halos located within the virial radius of a larger halo, as
additional halos, although we note that including them does not
significantly impact the particular results presented here.  In the
language of this paper, ``Milky Way mass halos'' simply refers to halos
the mass range $M_{\rm halo} = 10^{12 \pm 0.25} \; \Msun$. There are
$\sim 3.8 \times 10^5$ such halos at $z=0$ in our volume. 

We expect that the volume of our simulation should be sufficient to
statistically sample the largest \hii region sizes, which typically
grow to tens of comoving Mpc toward the end of the reionization epoch
\citep{furlanetto/oh:2005, zahn/etal:2005, mesinger/furlanetto:2007,
zahn/etal:2007, shin/etal:2008}.

\section{Results}

Having obtained a reionization time for each simulation particle, we
find that the global 68\% spread in reionization times across {\em
all} particles is $150$ Myr, and the spread in reionization times over
all present-day {\em halo} particles is 130 Myr. We begin in \S
\ref{sec:reionduration} by quantifying the internal scatter in
reionization times for halos of all masses.  In \S
\ref{sec:mwreionenv} we focus on Milky Way mass halos, identifying a
connection between halo reionization histories and local environment,
and in \S \ref{sec:virgom31}, we check how additional environmental
constraints, namely the presence of M31 and Virgo, affect our results.
Finally, in \S \ref{sec:varyingparams}, we briefly discuss how varying
the parameters of our reionization model affects the distribution of
Milky Way reionization histories. 

\subsection{Halo Reionization Histories and Durations} \label{sec:reionduration}

\begin{figure}[!t]
		\centering
		\includegraphics[width=0.99\colwidth]{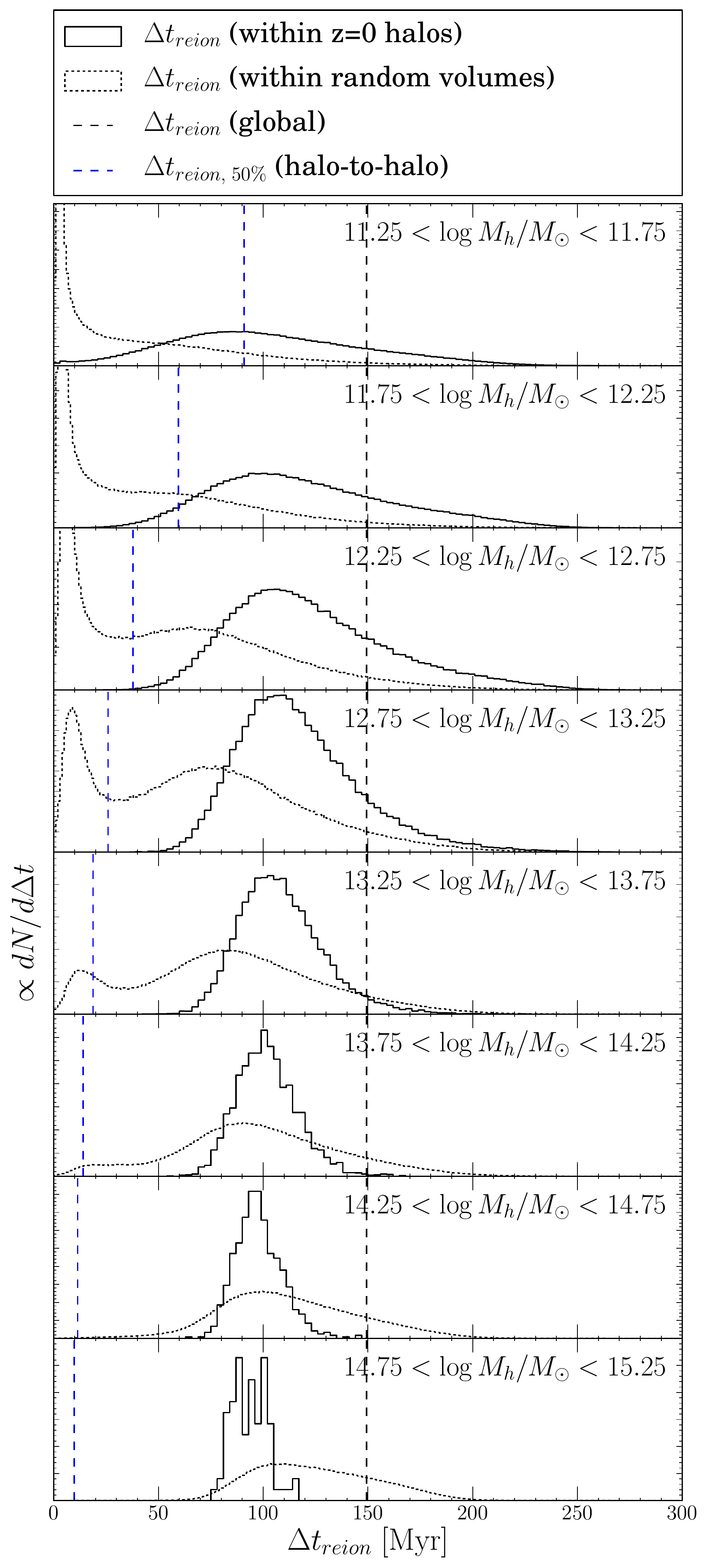}
		\caption{Distribution of $\Delta \treion$ within $z=0$
                  halos, for our fiducial model, in increasing mass
                  bins.  Plotted for comparison are the halo-to-halo
                  scatter in median $\zreion$ (blue) and the global
                  scatter for the all particles (black).  The dotted
                  curve represents the distribution of $\Delta
                  \treion$ for random cubic volumes of comparable
                  mass.} 
		\label{fig:treionspread}
		\vspace{0.75cm}
\end{figure}

Having tagged each simulation particle with a reionization redshift,
we obtain reionization histories for the mass in each $z=0$ halo.
Figure \ref{fig:haloreionhistories} shows the reionized mass fraction
of present-day halos as a function of redshift, for a random selection
in increasing mass bins: from $\sim 10^{11.5} \; \Msun$ up to $\sim
10^{15} \; \Msun$. For comparison, we also indicate the mean reionized
fraction of halos within each mass bin as well as the global reionized
mass fraction.  We note that most halos have biased reionization
histories, in that they reionize before the rest of the universe. The
majority of $z=0$ halos have a reionized fraction above the global
fraction at any given redshift, indicating that the halos reionize
early relative to the universe as a whole, or equivalently that they
are spatially correlated with regions of early reionization. In Figure
\ref{fig:halosreionizefirst}, we show this more explicitly by
displaying the redshift differences between halo and global
half-reionization. By this measure, most halos larger than $\sim
10^{12}\;\Msun$ and nearly all halos larger than $\sim
10^{12.5}\;\Msun$ reionized ahead of the universe. Note that the mean
halo reionization fractions in Figure \ref{fig:haloreionhistories} are
calculated at fixed redshifts, while Figure
\ref{fig:halosreionizefirst} compares halos at a fixed reionized
fraction, accounting for any apparent discrepancies between them.

For each $z=0$ halo, we calculate $\Delta \treion$, which we define as
the central 68\% spread in reionization times of its present-day
particles, i.e. the difference between the 16th and 84th percentile of
$\treion$ values within the halo.  Thus, the quantity $\Delta \treion$
quantifies the $1\sigma$ scatter in mass-weighted reionization times
within a present day halo, and it may be thought of as approximately
characterizing the duration of a halo's reionization history.
However, we caution against an overly simplistic picture of the
process, noting that these halos were generally built up through the
mergers of many smaller halos, which were themselves distinct or
yet-unformed objects during the epoch of reionization. 

\begin{figure}[!t]
	\centering
	\includegraphics[width=\colwidth]{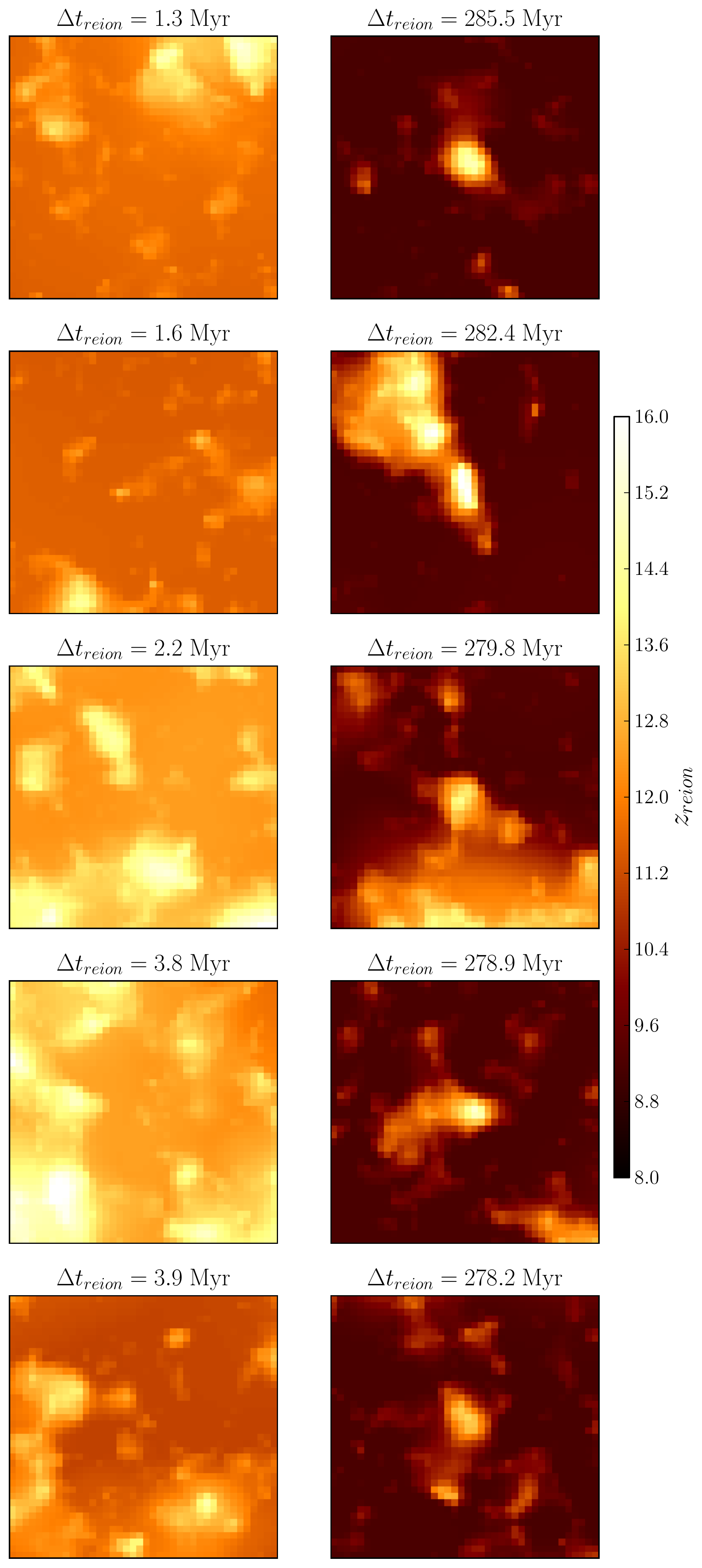}
	\caption{Slices through the 3D reionization map, centered on
          the \emph{initial} center of mass of particles in eventual
		  $z=0$ Milky Way mass halos ($10^{12\pm0.25} \;\Msun$).
          Plotted are mean-value projections, through a slice with
          co-moving dimensions: $12.3\times 12.3 \times 3.3$ $\Mpcph$.
          Note that each cell coincides with the initial, unperturbed
          position of a particle.  Reionization environments are shown
          for halos with the (\textbf{left}) shortest and
          (\textbf{right}) longest $\Delta \treion$.} 
	\label{fig:zrcubeslices}
\end{figure}

Figure \ref{fig:treionspread} shows the distribution of $\Delta
\treion$ for all halos in our fiducial reionization model, also in
logarithmically spaced mass bins, from $\sim 10^{11.5} \;\Msun$ up to
$\sim 10^{15}\;\Msun$.  We find that the median $\Delta \treion$ is
about $100$ Myr for $\sim 10^{11.5}\;\Msun$ halos, rising to $115$ Myr
for $\sim 10^{12} \; \Msun$ halos, then decreasing weakly with halo
mass to $95$ Myr for $\sim 10^{15} \; \Msun$ halos.  For comparison,
the 68\% scatter in median reionization times \emph{across} halos is
plotted in each mass bin, decreasing from $\sim 90$ Myr to $\sim 10$
Myr over the same mass range.  Significantly, $\Delta \treion$, the
scatter in $\treion$ \emph{within} halos, is typically larger than the
same scatter in median $\treion$ across halos: reionization epochs
within present-day halos are typically more variable within, rather
than across, halos of similar mass. 

For further comparison, Figure \ref{fig:treionspread} also shows the
distribution $\Delta \treion$ for completely \emph{random} cubic
volumes with equivalent masses, via the dotted curve.  We note that
``small'' random volumes, roughly at or below Milky Way masses, exhibit a
sharp excess of $\Delta \treion \lesssim 10$ Myr.  This indicates that
most random $\lesssim 10^{12}\;\Msun$ regions are reionized essentially
uniformly and likely have relatively little reionization structure.
Going to higher mass bins, this sharp peak in the distribution slowly
widens and essentially disappears in the $\sim 10^{14}\;\Msun$ bin,
giving way to a broader distribution of more extended $\Delta
\treion$.  The fact that there is still a fairly broad spread of
$\Delta \treion$ for random cluster-mass volumes implies that
reionization histories and reionization structure are quite variable
from region to region, even on these large scales. 

While the mean $\Delta \treion$ within a given halo mass bin does not
appear to be a strong function of halo mass, if at all, we
nevertheless find that smaller halos have a broader range of $\Delta
\treion$, implying a marked halo-to-halo diversity in reionization
histories and scenarios.  In particular, Milky Way mass halos exhibit
$\Delta \treion$ of anywhere from $\sim 1$ Myr to $\sim 300$ Myr. In
the next section, we specifically focus on the connection between
$\Delta \treion$ and environment for $10^{12 \pm 0.25} \; \Msun$
halos. 

\subsection{Reionization and Large Scale Environment} \label{sec:mwreionenv}

\begin{figure}[!t]
	\centering
	\includegraphics[width=\colwidth]{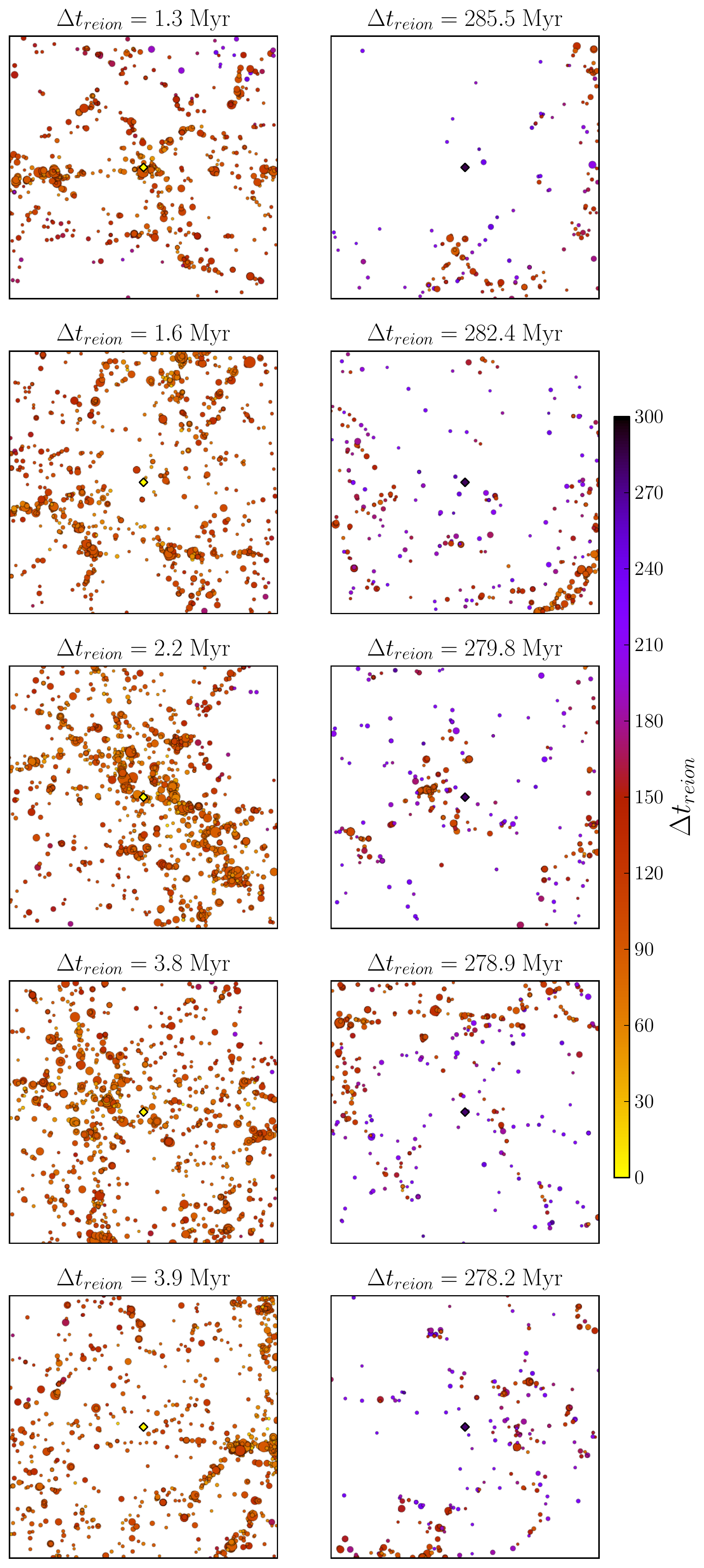}
	\caption{Present day surroundings of MW-mass halos, with the
		(\textbf{left}) 5 shortest and (\textbf{right}) 5 longest
		$\Delta \treion$.  Each plot represents a 2D projection of a
		rectangular volume, $50\times 50\times 20$ $\Mpcph$, centered
		around the $z=0$ halo of note (\emph{diamond}).  Surrounding
		halos (\emph{circles}) are themselves colored by their $\Delta
		\treion$. Only halos with mass $> 10^{11.75}\; \Msun$ are
		shown. Note: circle sizes scale monotonically with halo mass and
		do not directly represent physical halo size.}
	\label{fig:haloposslices}
\end{figure}

Figure \ref{fig:zrcubeslices} shows localized slices through the
reionization cube, centered on the regions which collapse into
present-day Milky Way mass halos, selected to show those with the
longest and shortest $\Delta \treion$.  MW halos with the shortest
$\Delta \treion$ form in regions which ionize slightly earlier
overall, whereas MW halos with the longest $\Delta \treion$ are
surrounded by regions which do not ionize until quite late.
Furthermore, Figure \ref{fig:haloposslices} shows the same 10 MW halos
in the context of their \emph{present-day} surroundings, with each
nearby halo marked with its own $\Delta \treion$.  While these slices
show only the most extreme cases, they suggest that halos with the
shortest $\Delta \treion$ are found in more clustered environments,
and are themselves surrounded by halos with relatively short $\Delta
\treion$.  By contrast, halos with the longest $\Delta \treion$ are
apparently found in relative isolation, and the few halos which
surround them also tend to have relatively long $\Delta \treion$.

Figure \ref{fig:trspread_v_env} displays the distribution of $\Delta
\treion$ for MW halos, binned by the abundance of nearby halos
(specifically, nearby halos larger than the cutoff $10^{11.75}
\;\Msun$, within 20 Mpc, at $z=0$).  The median $\Delta \treion$
ranges from $\sim 200$ Myr for halos in highly isolated environments
to $\sim 75$ Myr for halos in highly clustered environments.

That MW halos with short $\Delta \treion$ are preferentially found in
clustered environments suggests a physical picture for those halos'
reionization processes: namely, that MW halos with short $\Delta
\treion$ contain more ``externally'' reionized matter, which was
ionized by radiation from sources outside of the halo progenitors.
Within the context of our reionization model, we define ``internally''
and ``externally'' reionized particles following a criterion proposed
by \citet{alvarez/etal:2009b}.  We are able to assign each halo
particle two characteristic radii:
\begin{enumerate}
	\item $R_{\rm bubble}$, the radius of the region at which it first crossed the ionization threshold, and
	\item $R_{\rm Lag}$, the $z=0$ halo's Lagrangian radius, defined by $M_{\rm halo} = 4\pi\bar{\rho}\, R_{\rm Lag}^3/3$.
\end{enumerate}
We consider the particle internally reionized if $R_{\rm bubble} <
R_{\rm Lag}$, and externally reionized if $R_{\rm bubble} > R_{\rm
Lag}$.  We can thus obtain a halo's internally reionized mass
fraction, $f_{internal}$.

Figure \ref{fig:trspread_v_finternal} shows the correlation between
$f_{internal}$ and $\Delta\treion$, as well as between $f_{internal}$
and the abundance of nearby ($<20$ Mpc) halos (rank correlation
coefficients: $\rho=0.64$ and $\rho=-0.38$, respectively). Halos with
short reionization histories preferentially contain more externally
reionized mass, and more externally reionized halos tend to be found
in clustered environments.

\subsection{Milky Way Environmental Constraints: M31 and Virgo} \label{sec:virgom31}

\begin{figure}[!t]
		\centering
		\includegraphics[width=0.95\colwidth]{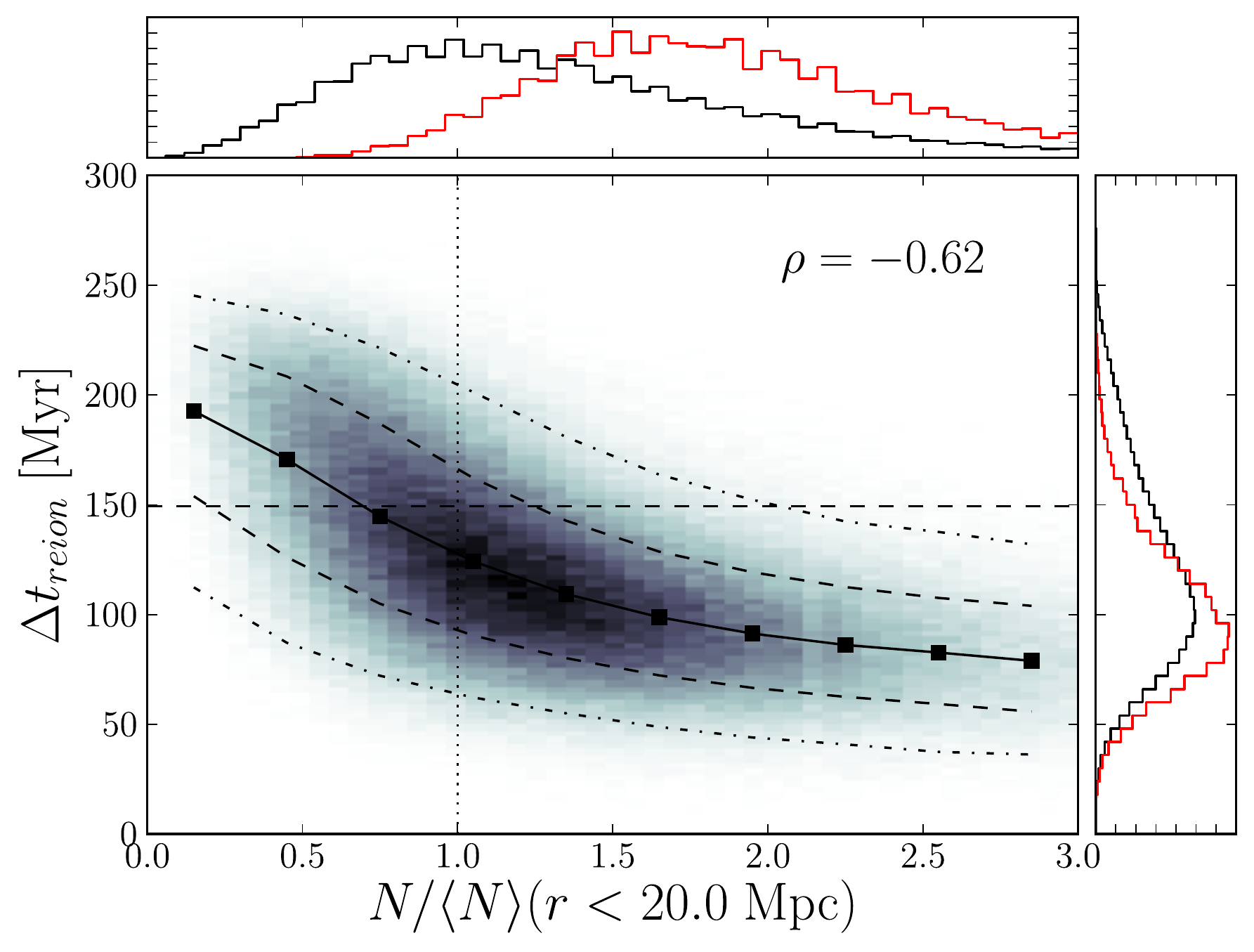}
		\caption{2D histogram of $\Delta \treion$ vs. the overdensity
			of neighboring halos (within 20 Mpc), for present-day
			Milky Way mass halos. Note that only $10^{12\pm 0.25}
			\;\Msun$ halos are histogrammed, but any halo larger than
			$10^{11.75} \;\Msun$ can be counted as one of their
			neighboring halos. Lines plot the median trend ({\em
			solid}), as well as the 1$\sigma$ ({\em dashed}) and
			2$\sigma$ ({\em dash-dotted}) spread. The Spearman rank
			correlation coefficient $\rho$ is also indicated. The
			horizontal dashed line represents $\Delta \treion$ for the
			entire box, while the vertical dotted line explicitly
			marks the mean number density of halos. Individual 1D
			distributions are also shown, both for all MW mass halos
			(\emph{black}) and those located near a Virgo mass halo
			(\emph{red}, see Section \ref{sec:virgom31}).}
		\label{fig:trspread_v_env}
\end{figure}

\begin{figure}[!t]
		\centering
		\includegraphics[width=0.95\colwidth]{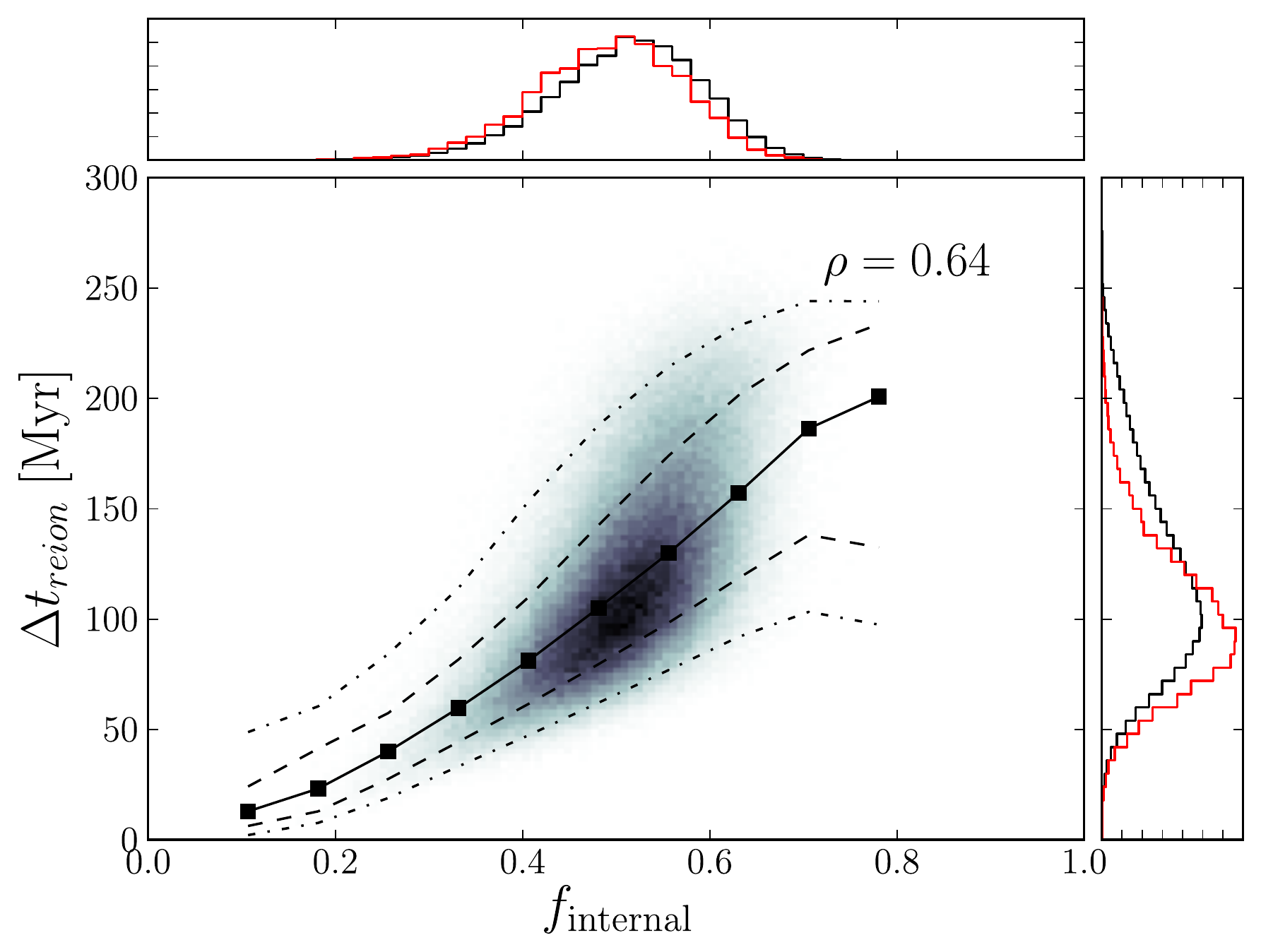}
		\includegraphics[width=0.95\colwidth]{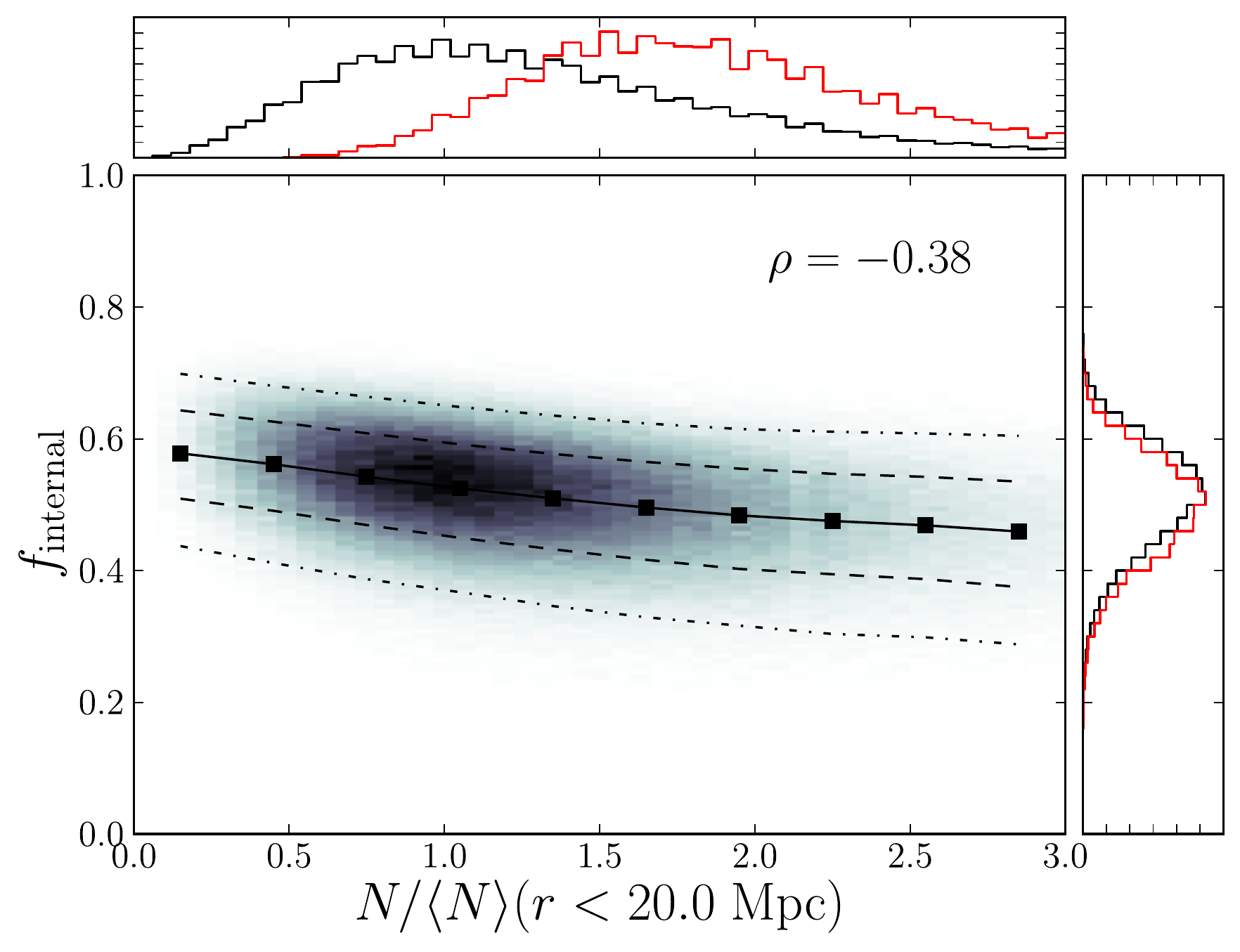}
		\caption{Correlations between $f_{internal}$ and both
		variables from Figure \ref{fig:trspread_v_env}
		($\Delta\treion$ and local halo number density). See that
		figure for more details. Individual 1D distributions are also
		shown, both for all MW mass halos (\emph{black}) and those
		located near a Virgo mass halo (\emph{red}, see Section
		\ref{sec:virgom31}). \textbf{Top}: 2D histogram of $\Delta
		\treion$ vs. $f_{internal}$, the fraction of ``internally''
		reionized mass for present-day MW halos. \textbf{Bottom}: 2D
		histogram of $f_{internal}$ vs. the number density of halos
		within 20 Mpc, for present-day MW halos.}
	\label{fig:trspread_v_finternal}
	\label{fig:finternal_v_env}
\end{figure}

Naturally, one wishes to determine where to place the true Milky Way
halo in the aformentioned distributions. There exist numerous previous
studies which have selected Milky Way candidates from simulations by
matching, for example, the observed mass, velocity, and distance
properties of the Local Group, the presence of the Virgo Cluster,
and/or the local number density of observable galaxies
\citep[e.g.,][]{governato/etal:1997, forero-romero/etal:2011,
schlegel/etal:1994, few/etal:2012}. 
A comprehensive investigation of the connection between the true Local
Group environment and its reionization history is warranted but is
left for future study. As an exploratory check, we select out
candidates based on distance constraints to M31 and the Virgo Cluster.
The following criteria were separately used:
\begin{enumerate}
	\item MW mass halos $750 \pm 50$ kpc away from exactly one other $10^{12 \pm 0.25}\;\Msun$ halo, corresponding to a MW-M31 distance constraint. \label{m31constraint}
	\item MW mass halos $16.5 \pm 1.1$ Mpc away from exactly one $\sim 10^{14.4 \pm 0.25} \Msun$ halo, corresponding to a MW-Virgo Cluster constraint. \label{virgoconstraint}
	\item MW mass halos which satisfy \emph{both} of the previous constraints. \label{m31virgoconstraint}
\end{enumerate}
The distance constraints, while deliberately lenient, are consistent
with measured values from the literature
\citep[e.g.,][]{mcconnachie/etal:2005, mei/etal:2007}. From our full
sample of 383,237 Milky Way mass halos, the M31-only constraint
selected out 1.7\% of the sample, the Virgo-only constraint selected
about 7.0\%, and the combined M31-Virgo constraint selected out 0.12\%
(458 candidate halos).  Including M31 in our criteria (Constraints
\ref{m31constraint} and \ref{m31virgoconstraint}) did not
statistically change any of the distributions presented here. The
Virgo-only constraint (Constraint \ref{virgoconstraint}) does have a
noticeable effect, as indicated by the red 1-D distributions in
Figures \ref{fig:trspread_v_env} and \ref{fig:trspread_v_finternal}:
from the full sample, it selects out MW halos with a typically higher
local overdensity, shorter reionization history, and slightly lower
internally reionized mass fraction.

The masses of our Virgo candidates are consistent with a range of
measured values, $(1.4 - 4.2)\times 10^{14} \;\Msun$
\citep{ferrarese/etal:2012}, though we note that there is significant
uncertainty in these measurements. Increasing the mass of Virgo
candidates selects even more preferentially for MW halos that have
shorter reionization histories, are in more overdense environments,
and have lower $f_{internal}$.

The presence of Virgo is of particular interest since its progenitors
may have reionized some or much the present-day Local Group
\citep[e.g.][]{weinmann/etal:2007, alvarez/etal:2009b}. Since we find
that Milky Way halos near a Virgo-like cluster typically have slightly
more externally reionized matter, it is plausible that Virgo was
responsible for externally reionizing at least some material in the
Milky Way.

\subsection{Varying Reionization Model Parameters} \label{sec:varyingparams}

\begin{figure*}[!t]
		\centering
		\includegraphics[width=0.93\textwidth]{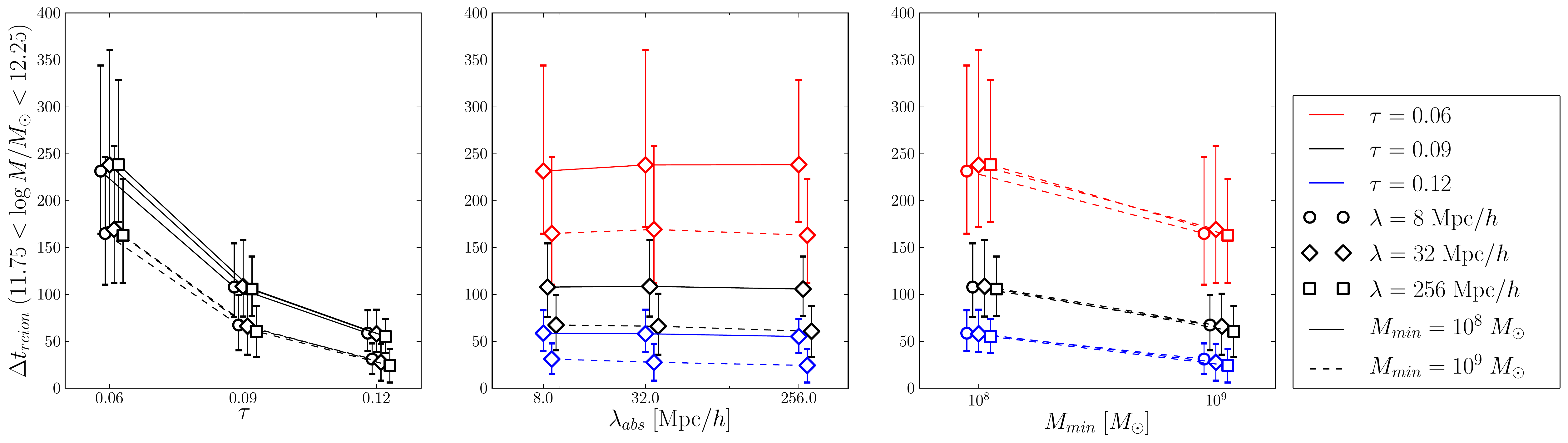}
		\caption{Values of $\Delta \treion$ for Milky Way mass halos,
			over different reionization parameter values.  Median
			values (points) and 1$\sigma$ spread (error bars) are
			shown for each combination of $\taues$, $\labs$, and
			$\Mmin$.  In each panel, a different parameter is
			independently varied on the horizontal axis, but each
			panel contains the same 18 points and error bars.  Point
			have been plotted with a slight horizontal offset for
			readability.  \textbf{Left}: Variation of $\Delta \treion$
			with $\taues$, the optical depth to reionization.  In
			general, higher values of $\taues$ correspond to earlier
			reionization scenarios.  \textbf{Middle}: Variation of
			$\Delta \treion$ with $\labs$, the mean free path of Lyman
			absorption systems.  \textbf{Right}: Variation of $\Delta \treion$ 
			with $\Mmin$, the minimum halo mass of ionizing sources.}
		\label{fig:varyparams}
		\vspace{0.5cm}
\end{figure*}
Figure \ref{fig:varyparams} displays the values of $\Delta \treion$
obtained for Milky Way mass halos, over all combinations of parameter
values in our reionization model.  Holding all other parameters fixed,
decreasing $\taues$ to 0.06 increases the median MW $\Delta \treion$
to $\sim 240$ Myr, and increasing $\taues$ to 0.12 decreases the
median MW $\Delta \treion$ to $\sim 60$ Myr. The strong decrease in
reionization duration with increasing $\taues$ is due mainly to the
shorter Hubble time for earlier reionization. Increasing $\Mmin$
independently to $10^9 \;\Msun$ also shortens $\Delta \treion$ in MW
halos to $\sim 70$ Myr.  This is because higher mass halos are rarer,
and therefore grow in abundance more quickly at a given redshift,
leading to a shorter reionization duration. 

Varying $\labs$ over the values $(8,32,256) \;\Mpcph$ appears to have
little to no systematic effect on the 1$\sigma$ spread of $\Delta
\treion$ values for MW halos. This is expected, mainly for two
reasons. First, as shown by \cite{alvarez/abel:2012}, the absorption
system mean free path affects the final percolation phase of
reionization, when the global neutral fraction is less than $\sim
0.1$.  Given that our definition of the duration of reionization is
the range over which $0.16<x<0.84$, on average the duration of
reionization within individual halos should not change much. Second,
while smaller $\labs$ introduces more inhomogeneous reionization
structure on the scale of $\sim \labs$, even scales of $\labs \sim 8
\; \Mpcph$ are still well outside the $\sim 1 \; \Mpcph$ Lagrangian
radii of $10^{12} \;\Msun$ halos. Those MW halos that are reionized
internally therefore are not sensitive to changes in the mean free
path $\labs$. Further work will be required to assess the effect of
varying $\labs$ on externally reionized halos.

We find that the inverse correlation of $\Delta \treion$ with
abundance of nearby halos is present in all models.  However, in
scenarios that combine high values of $\taues$ and $\Mmin$ ($\taues =
0.09,\; 0.12$ and $\Mmin=10^9\;\Msun$), we find many MW mass halos
with $\Delta \treion \approx 0$. This indicates a significant
population of such halos with highly homogeneous reionization
histories, for these particular reionization scenarios. Such models,
however, are at the extreme end of our parameter space and are
disfavoured -- if $\taues>0.09$, it is more likely that the
reionization history is more complex than the models presented here,
perhaps due to photoionization feedback
\citep[e.g.,]{iliev/etal:2007,alvarez/etal:2012}, substantial early
ionization by massive stars in minihalos
\citep[e.g.,][]{abel/etal:2007,ahn/etal:2012}, or even
pre-reionization by x-rays
\citep[e.g.,][]{ricotti/ostriker:2004,ricotti/etal:2005,mcquinn:2012}.

\section{Discussion}

We have simulated the reionization of a 600 Mpc box and characterized
reionization histories of $z=0$ halos, focusing on Milky Way mass
halos and the dependence of $\Delta \treion$ on environment. Our main
results are as follows:

\begin{enumerate}
	\item We quantify the typical spread in reionization epochs within $z=0$ halos.  In our fiducial reionization model, we find that in our smallest halos, $\sim 10^{11.5}\;\Msun$, the median scatter in reionization times is $\Delta\treion\sim 100$ Myr. In Milky Way mass halos, $\sim 10^{12}\;\Msun$, this increases slightly to $\Delta \treion \sim 115$ Myr.  For the largest cluster mass halos, $\sim 10^{15} \;\Msun$, this decreases slightly to $\Delta \treion \sim 95$ Myr.

	\item The typical scatter of reionization epochs \emph{within halos} is notably larger than the scatter \emph{across halos}, provided the ``reionization epoch'' of a single halo is defined by its median reionization time.  This is true for all halo masses considered in this study.  To be explicit, the scatter \emph{across} halos is $\sim 90$ Myr for $\sim 10^{11.5} \;\Msun$ halos, decreasing to $\sim 10$ Myr for $\sim 10^{15}\;\Msun$ halos.

	\item For MW mass halos, the halo-to-halo diversity in $\Delta \treion$ is significant (ranging at the extremes from $\sim 1$ Myr to $\sim 286$ Myr).  The values of $\Delta \treion$ for individual halos correlate with both their reionization environments and present-day environments.  MW mass halos with short (long) $\Delta \treion$ were formed from regions which reionized relatively homogeneously (inhomogeneously) and slightly earlier (later).  They also tend to be found in clustered (isolated) present-day environments, i.e. they are surrounded by relatively more (fewer) halos within a radius of 20 Mpc.  A larger fraction of their mass is likely to have been externally (internally) reionized.

	\item Varying the global parameters of the reionization model will affect the $\Delta\treion$ values in MW mass halos. Increasing $\taues$ decreases the typical $\Delta \treion$.  Increasing $\Mmin$ also decreases the typical $\Delta \treion$.  Varying $\labs$ has little to no systematic effect on the typical $\Delta \treion$ values of MW mass halos.
\end{enumerate}

Our results suggest that the reionization histories of present-day
halos are inadequately characterized by a single $\zreion$, and this
has potential implications for modeling the satellite galaxies of the
Milky Way. For simplicity, a single $\zreion$ has often been assumed
in galaxy formation models \citep[e.g.][]{koposov/etal:2009,
busha/etal:2010}. Under such an assumption, \cite{busha/etal:2010}
found that the timing of the reionization epoch may strongly affect
the present-day Milky Way satellite population, with the number
varying by up to two orders of magnitude over the reionization epochs
$\zreion\approx 6-12$. However, using a single $\zreion$ per halo
implicitly assumes that scatter of reionization times within halos is
smaller than the scatter across halos. Our results suggest otherwise:
the spread of reionization times within halos is \emph{not} negligible
compared to the spread of (median) reionization times across halos.
This result would thus reduce the predicted halo-to-halo variation in
satellite populations across Milky Way mass halos.

In other words, $\Delta\treion$ is essential to fully characterize the
impact of reionization on the $z=0$ halo, with the minority exception
of halos that were reionized rapidly, likely by external sources. We
did find that in scenarios combining high $\taues$ (0.09, 0.12) and
$\Mmin$ ($10^9\;\Msun$), there were many MW mass halos with short
$\Delta \treion$, as compared to the halo-to-halo scatter.  In these
scenarios---corresponding to relatively early reionization by rare but
highly efficient ionizing sources---many MW mass halos would be better
characterized by a single reionization epoch, $\zreion$, than by a
spread, $\Delta\treion$.

Notably, we find a correlation between the reionization history of
Milky Way halos and their environment, as well as their fraction of
internally reionized mass (Figures \ref{fig:trspread_v_env} and
\ref{fig:finternal_v_env}). This contrasts somewhat with the results
of \cite{weinmann/etal:2007}, which found no statistical correlation
between reionization histories of field halos---either in terms of
reionization epoch or the number of externally reionized halos---and
their environment.  However, it is worth pointing out that our
definitions differ from theirs. Rather than a binary definition of an
internally vs. externally reionized halo, we defined an internally
reionized mass fraction for each halo, which does correlate with
reionization history and environment.  Our criteria for internally vs.
externally reionized mass is also fundamentally different, owing
partly to a difference in reionization model.  Additionally, we have
characterized halo reionization histories differently, noting that
most are more accurately characterized by $\Delta \treion$ rather than
a single epoch $\zreion$.

The values of $\Delta \treion$ found here are likely to be lower
limits. This is because our model does not self-consistently include
or treat photoheating feedback in galaxy formation, which could act as
a self-regulation mechanism by suppressing ionizing radiation released
by halos in already-reionized regions.  Additionally, in this model,
once a cell crosses the reionization threshold, it remains ionized:
this would not account for so-called relic \hii regions, which might
form around short-lived early sources \citep[e.g.,][]{wise/abel:2008},
or other multiply reionized regions.  We also note that $\labs$ in
this model is, for simplicity, both spatially uniform and constant in
time, while in reality it is likely to be neither. We leave fully
self-consistent treatments of absorption systems to future studies,
although we reiterate that varying $\labs$ did not significantly
affect the specific findings we presented here.

Observationally, the reionization history of the Milky Way might be
imprinted in its satellite population. For example, if star formation
histories in at least some ultra-faint dwarfs were suddenly truncated
during reionization \citep{brown/etal:2012}, and one obtained a
statistical sample of such satellites, then the spread in their
truncation epochs should correspond to $\Delta\treion$ of the Milky
Way. However, such truncation epochs would need to be determined at a
precision of $\lesssim$ 50 Myr, well below the current $\sim$ Gyr
uncertainties quoted in \cite{brown/etal:2012}.  

The local reionization history could also affect the number of faint
satellite galaxies. As previously noted, earlier reionization could
suppress star formation and decrease the number of faint satellites,
but the effect of varying $\Delta \treion$ on any individual halo is
unclear without further knowledge of environment and assembly
history. When comparing instantaneous and inhomogeneous reionization
scenarios, \cite{lunnan/etal:2012} found no systematic offset in
$N_{\rm sat} \,(M_V \gtrsim -10)$, while acknowledging their small
sample size of 6 Milky Way halos.  Here, we have characterized the
reionization histories of a large sample of Milky Way mass halos, at
the expense of resolving their internal substructures.  Further study,
retaining a statistically significant sample of halos while resolving
their subhalos, will be necessary to fully understand the impact of
patchy reionization on the satellite population of Milky Way halos. 

\acknowledgments TYL is supported by a National Science Foundation
Graduate Research Fellowship.  This work received support from the
National Science Foundation under grant NSF-AST-0908883. We thank
Michael Busha for several useful discussions, and, along with the rest
of the LasDamas collaboration for providing access to the Las Damas
simulations, which were run on the Orange cluster at SLAC and on the
NSF TeraGrid machine Ranger (PI: Andreas Berlind). The reionization
simulations were performed in part on the GPC supercomputer at the
SciNet HPC Consortium. SciNet is funded by: the Canada Foundation for
Innovation under the auspices of Compute Canada, the Government of
Ontario, Ontario Research Fund--Research Excellence, and the
University of Toronto. We also thank Marla Geha for helpful
discussions and comments.

\bibliographystyle{apj}
\bibliography{mwreion}

\appendix
\begin{figure*}[!t]
	\centering
	\includegraphics[width=0.43\textwidth]{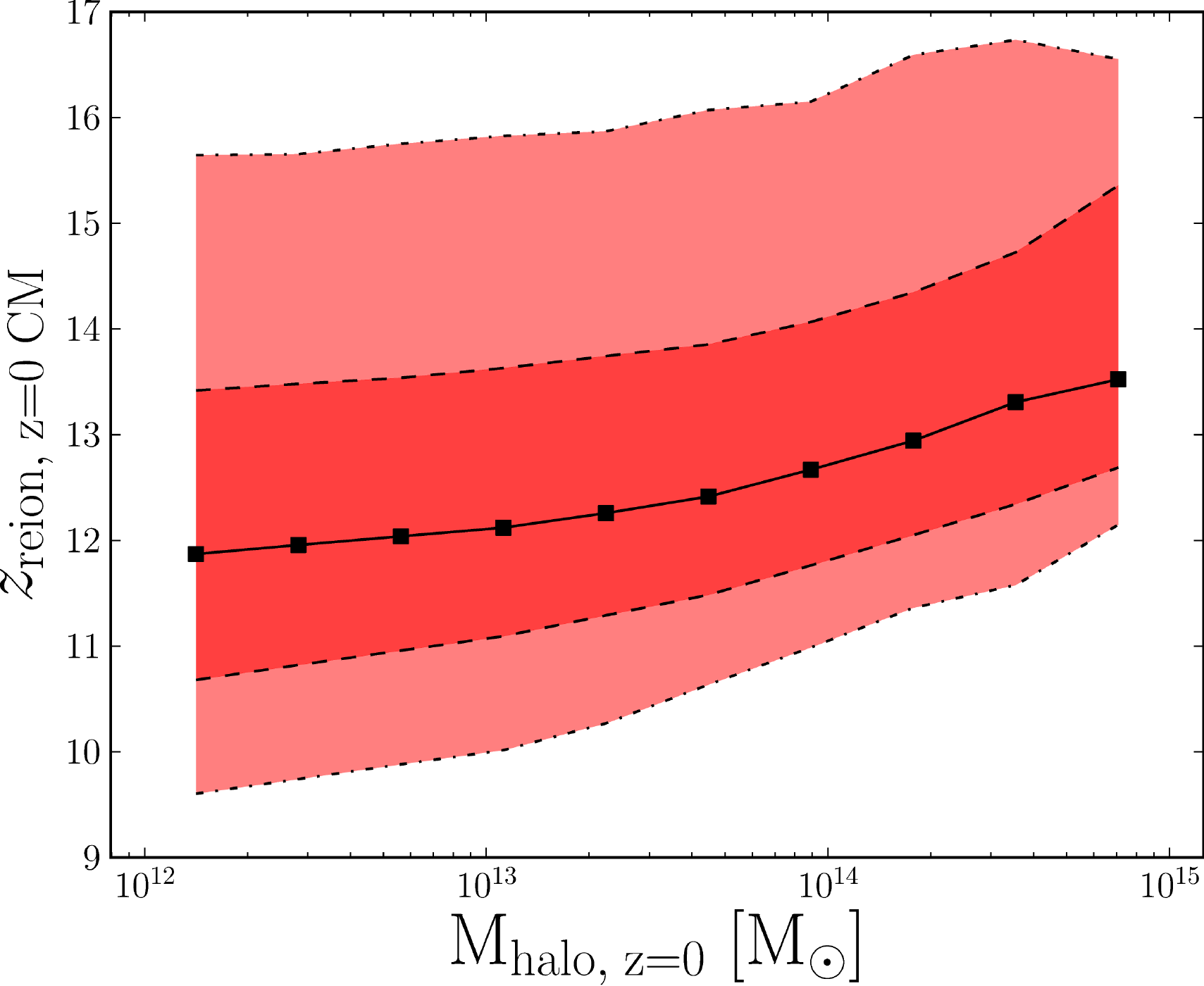}
	\hspace{1.5cm}
	\includegraphics[width=0.43\textwidth]{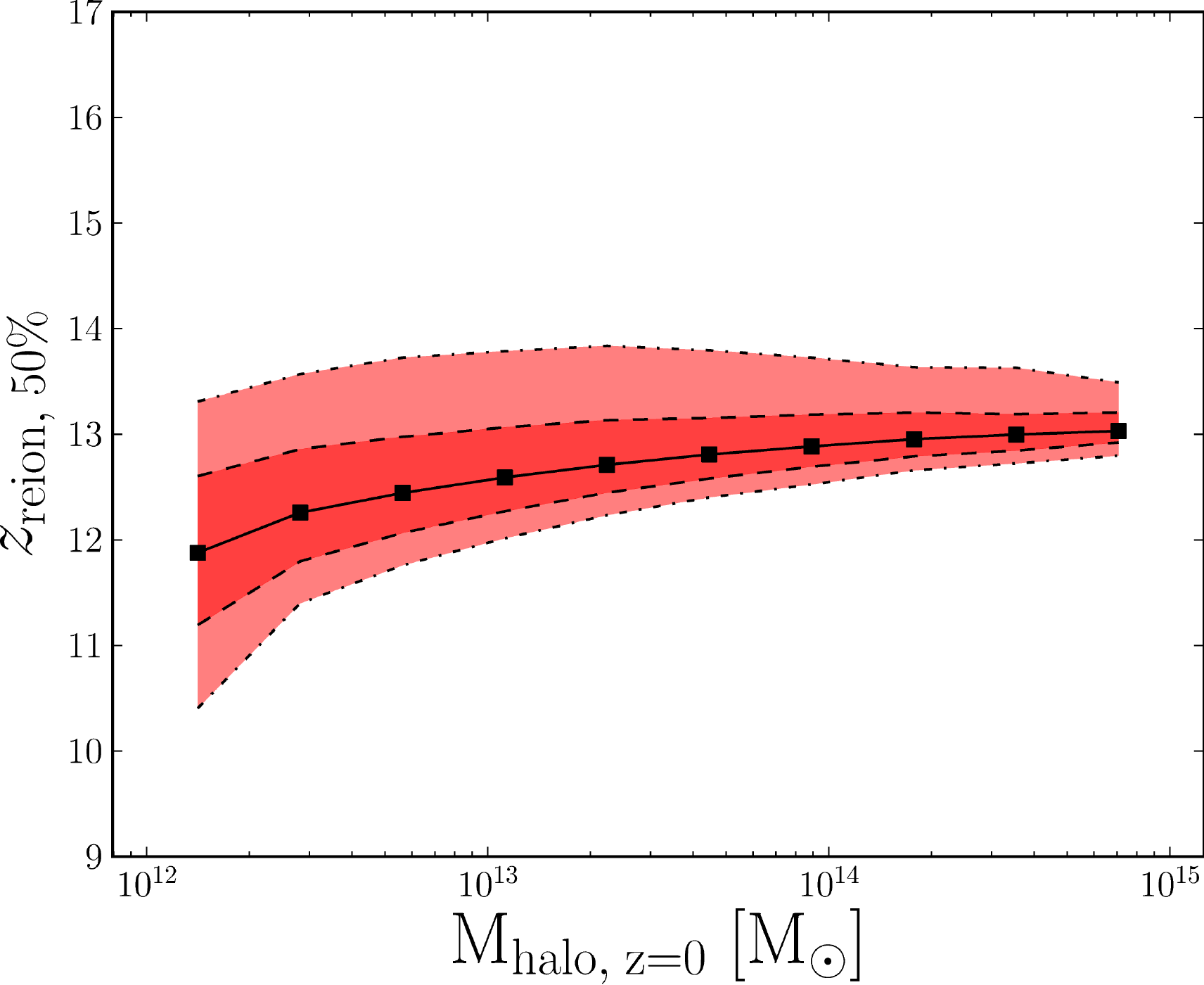}
	\caption{\textbf{Left}: Halo reionization redshifts, as a function
		of halo mass, using $\zreion$ of the $z=0$ halo center of mass.
		\textbf{Right}: Median halo reionization redshifts, as a function
		of $z=0$ halo mass.  Plotted in both are the median trend (solid
		black), as well as the 68\% (dashed) and 95\% (dot-dashed) 
		distributions.}
	\label{fig:zreion_v_mhalo}
\end{figure*}

\section{Halo Reionization Redshifts}\label{sec:halozreions}

Here we further discuss the assertion that halo reionization
histories, for masses considered in this study, are inadequately
characterized by a single reionization redshift.  In particular, we
expand on \cite{alvarez/etal:2009b}, in which each halo was assigned a
reionization redshift corresponding to the comoving cell around its
$z=0$ center of mass (CM).  This was justified by asserting that most
halos would not achieve sufficiently high peculiar velocities to
displace the halo CM out of the original ionizing bubble region.  The
left plot of Figure \ref{fig:zreion_v_mhalo} shows the distribution of
halo $\zreion$ in this study, by this definition, as a function of
mass. 

\begin{figure}[!t]
	\centering
	\includegraphics[width=0.33\textwidth]{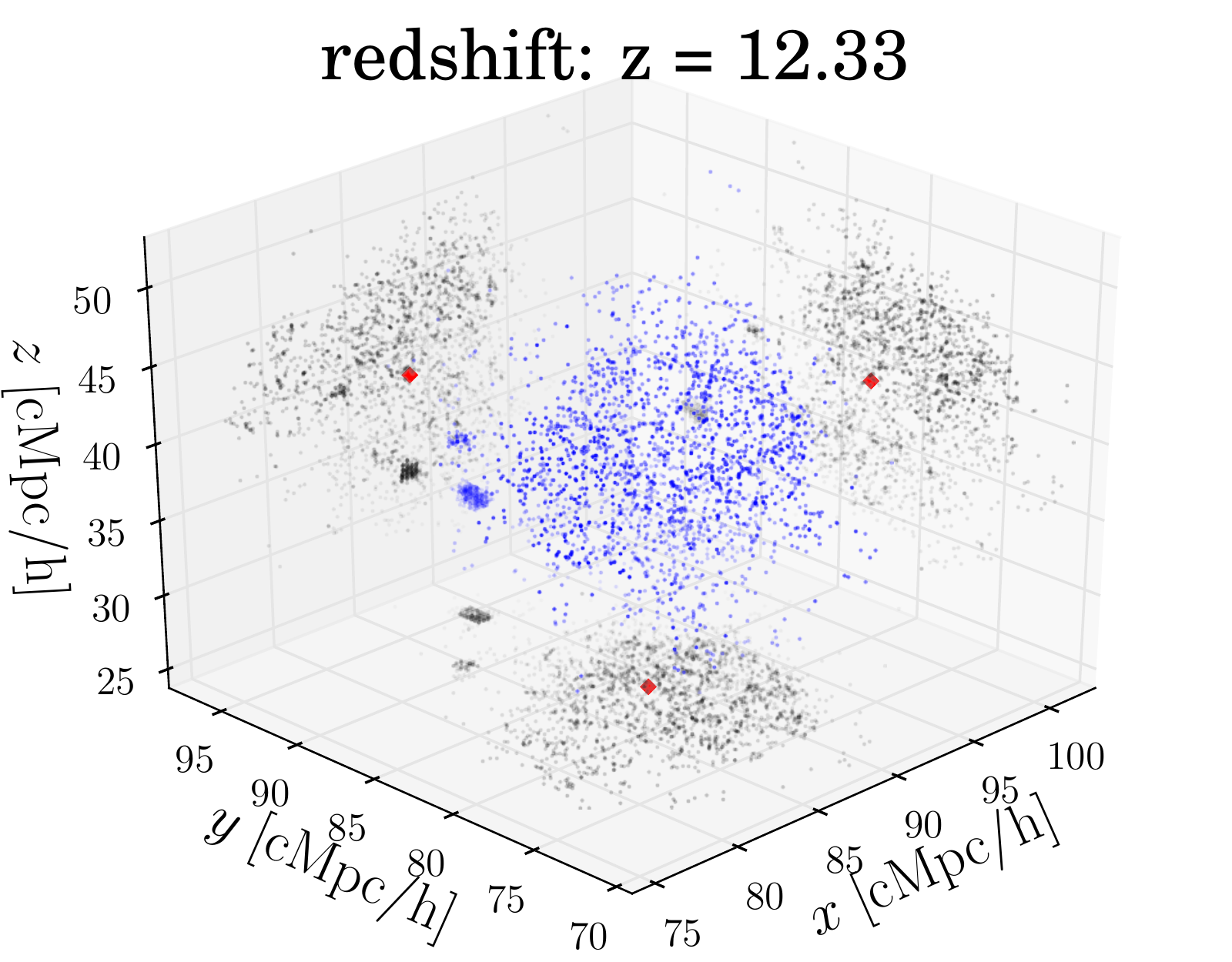}
	\includegraphics[width=0.33\textwidth]{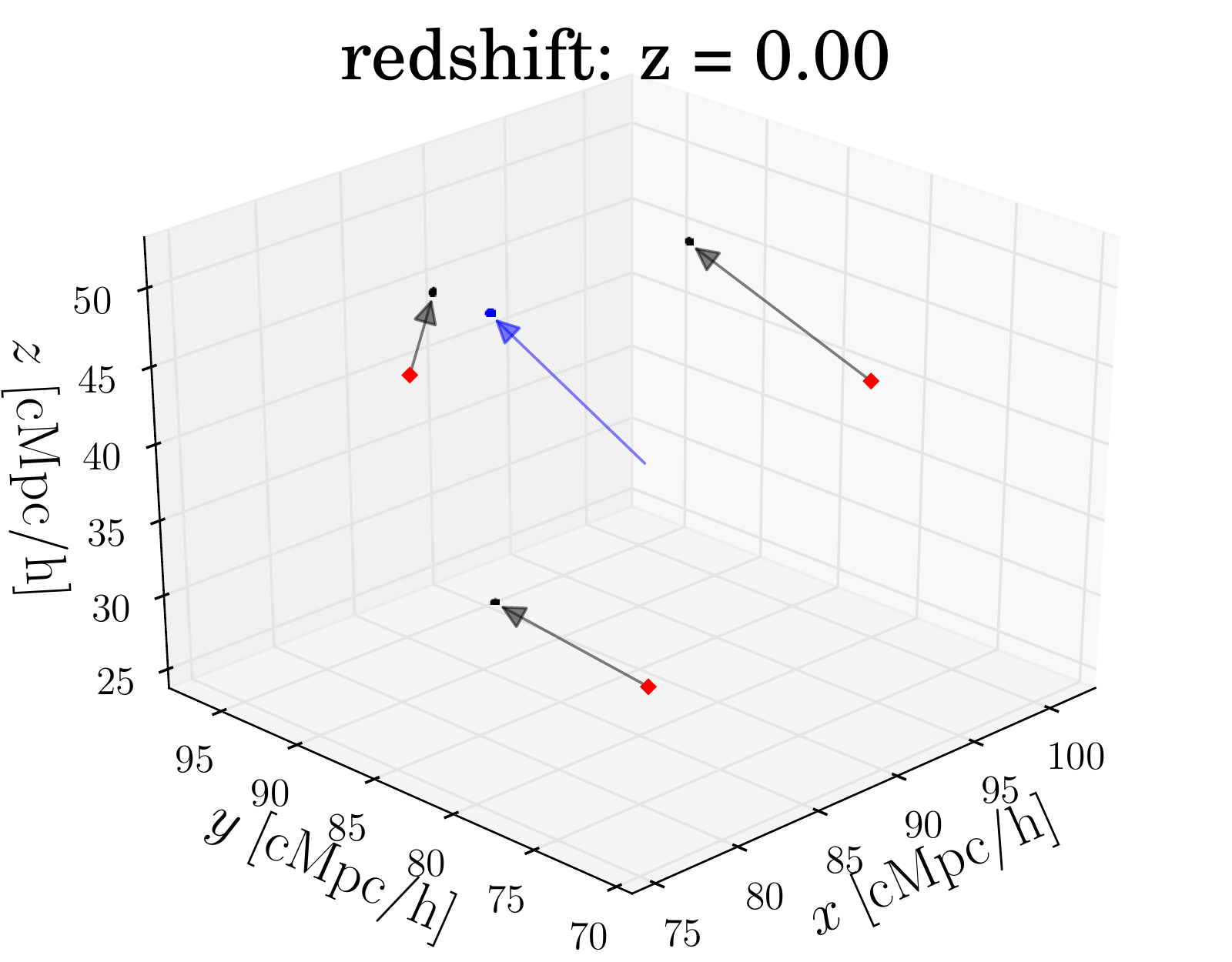}
	\includegraphics[width=0.33\textwidth]{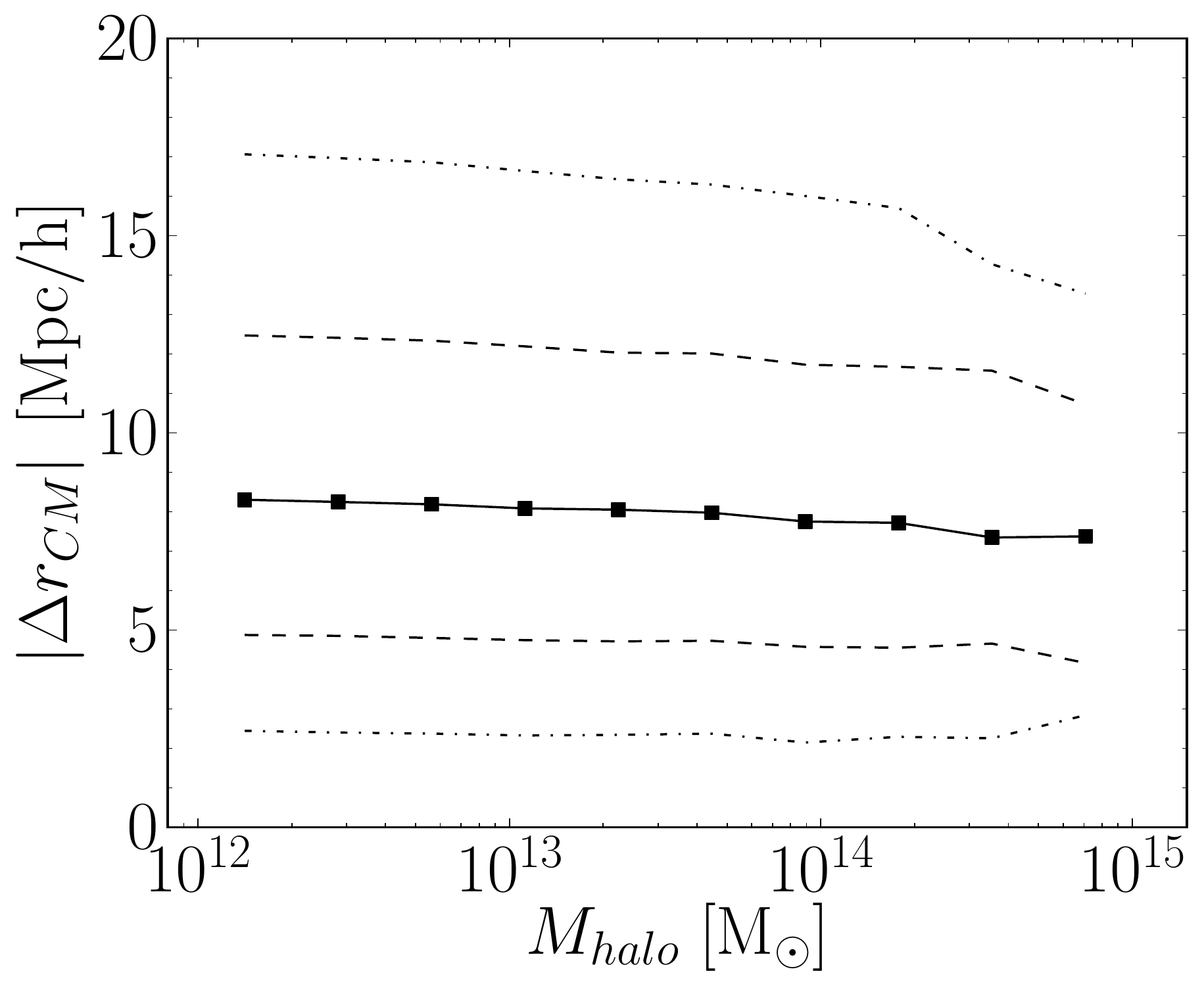}
	\caption{ \textbf{Left}: Particles of a randomly selected
		present-day halo, but in their positions at redshift $z=12.33$.
		Shown here are particle positions in 3D (\emph{blue points}),
		projections on the $xy$-, $yz$-, and $xz$- planes (\emph{gray
		points}), and the projected initial center of mass (\emph{red
		diamonds}).
		\textbf{Middle}: Same halo particles at redshift $z=0$, with
		the projected initial center of mass (\emph{red diamonds})
		retained for reference.  Arrows indicate the 3D (\emph{blue})
		and projected (\emph{gray}) total displacement of the center
		of mass.  \textbf{Right}: Total center-of-mass displacement of
		$z=0$ halos, as a function of mass.  The solid line follows
		the median in each mass bin, while the dashed and dash-dotted
		lines follow the 68\% and 95\% distributions, respectively.}
	\label{fig:cmdisp2}
\end{figure}

However, if the particles of all $z=0$ halos are tracked over the
course of the entire simulation, the last panel of Figure
\ref{fig:cmdisp2} shows the distribution of their total CM
displacements, after correcting for boundary periodicity, as a
function of mass.  From the plot, it is apparent that the distribution
is relatively consistent over all halo masses in this study.
Evidently, the typical comoving CM displacement is $\sim 8 \;\Mpcph$,
or $\sim 25$ cells in our reionization cube, but significantly larger
displacements are not uncommon.  Thus, while present-day massive halos
are preferentially found in overdense regions, which correlate with
earlier reionization redshifts, the $\zreion$ values thus obtained are
not necessarily indicative of the reionization redshift of matter
\emph{presently} in the halo, especially for smaller halos.  Even if
they are, such values of halo $\zreion$ are subject to scatter across
halos that is comparable to the scatter in $\zreion$ within halos, and
so it is generally unclear whether a single such $\zreion$ for an
entire halo characterizes the beginning, middle, or end of its
reionization history. 

A more consistent marker might be redshift at which the halo mass is
50\% reionized, $z_{reion, \; 50\%}$, but we find that this correlates
more weakly with halo mass in the mass range $10^{12} \; \Msun <
M_{halo} < 10^{15} \; \Msun$, as shown in the second panel of Figure
\ref{fig:zreion_v_mhalo}. We interpret this trend to
indicate that more massive present day halos are still generally
correlated with regions of earlier reionization.  However, we
interpret the weakness of the trend and the narrowness of the
distribution at high masses (approaching $\sim 10^{15} \; \Msun$) by
invoking hierarchical structure formation.  Present-day halos were
generally built up from small halos via mergers, many of which occured
after reionization.  While the reionization histories of distinct
halos at the end of reionization may vary greatly, $z=0$ halos
aggregate those once-separate reionization histories.  More massive
halos have aggregated more disparate reionization histories from a
much larger Lagrangian region, and so the halo-to-halo variation in
their $z_{reion,\; 50\%}$ values will be suppressed.

\end{document}